\documentclass[aps,prd,onecolumn,superscriptaddress,groupedaddress,nofootinbib,preprintnumbers]{revtex4}
\pdfoutput=1 \usepackage{amsmath,amssymb,pstricks,graphicx,todonotes,xspace}
\usepackage{pgfplots} \presetkeys{todonotes}{inline}{}
\newcommand{\professor}{\texttt{Professor}}
\newcommand{\rivet}{\texttt{Rivet}}
\newcommand{\herwig}{\texttt{Herwig\,7}}
\newcommand{\pythia}{\texttt{Pythia\,8}}
\newcommand{\sherpa}{\texttt{Sherpa}}
\newcommand{\thepeg}{\texttt{ThePEG}}
\usepackage{amsmath,amssymb,amsfonts,enumitem,mathtools}
\usepackage{cleveref} \usepackage{graphicx} \usepackage{subfigure}

\begin{document} \hfill LU-TP-19-40, MCNET-19-20

\title{High dimensional parameter tuning for event generators}

\author{Johannes Bellm}\email{johannes.bellm@thep.lu.se} 
\affiliation{Department of Astronomy and Theoretical Physics, Lund University, S-223 62 Lund, Sweden} 
\author{Leif Gellersen}\email{leif.gellersen@thep.lu.se}
\affiliation{Department of Astronomy and Theoretical Physics, Lund University, S-223 62 Lund, Sweden}

\begin{abstract} 
Monte Carlo Event Generators are important tools for the understanding of physics at 
particle colliders like the LHC. In order to best predict a wide variety of observables, 
the optimization of parameters in the Event Generators based on precision data 
is crucial. However, the simultaneous optimization of many parameters is 
computationally challenging.
We present an algorithm that allows to tune Monte Carlo Event
Generators for high dimensional parameter spaces. To achieve this we first
split the parameter space algorithmically in subspaces and perform a \professor{}
tuning on the subspaces with bin wise weights to enhance the influence of
relevant observables.  We test the algorithm in ideal conditions and in real
life examples including tuning of the event generators \herwig{}  and \pythia{}
for LEP observables.  Further, we tune parts of the \herwig{} event generator with
the Lund string model.  
\end{abstract}

\maketitle

\section{Introduction and Motivation} \label{sec:intro}

The amount of data taken at the LHC allows measuring observables that can be
calculated perturbatively to high precision. 
This is beneficial for the comparison as well as the 
improvement of phenomenologically motivated non-perturbative models and also 
the searches for new physics.  
With the increasing precision made available in recent years through perturbative higher
order calculations, theoretical uncertainties have reduced dramatically.  In the
comparison of these theory predictions and experimental data, Monte Carlo event
generators (MCEG) \citep{Buckley:2011ms} like \herwig{} 
\citep{Bahr:2008pv,Bellm:2015jjp,Platzer:2011bc,Bellm:2017bvx},
\sherpa{} \citep{Gleisberg:2003xi} or
\pythia{} \citep{Sjostrand:2006za,Sjostrand:2014zea} play an important role.  
If possible a matched calculation that includes the perturbative 
corrections and the effects described by the MCEG can give an improved picture of the event structure measured by the 
experiment. 

Here event generators typically include additional phenomenological 
models to include effects that are not part of specialised fixed 
order and resummed calculations. 
Uncertainties of these additional modelled, but factorised, parts 
of the simulation can be estimated from lower order simulations.
The MCEG contain various, usually factorized (e.g. by energy scales) components. 
In the development, these parts can be improved individually.  
The afore mentioned matching to perturbative calculations is an example of 
recombining parts usually separated in the event generation, 
namely the parton shower and hard matrix element calculation.  
While it is possible to make such modifications and improvements, it is also necessary to keep other parts of the
simulation in mind.  
Even though the generation is factorized, various parts of
the simulation will have an impact on other ingredients of the generator. 
Any modification can, in general, have an impact on the full
events. 
Calculated, or at least theoretically motived improvements will lead
to a reduction of freedom that eventually also restricts the parameter ranges
of the phenomenological models that could be used to compensate the variations
of the perturbative side 
\citep{Reichelt:2017hts,Gieseke:2017clv,Hoche:2017iem,Hoche:2017hno,Bellm:2018wwz,Duncan:2018gfk,Dulat:2018vuy,Bewick:2019rbu}. 
The capability to describe data needs to be
reviewed with the modifications made in order to use the event generator for
future predictions or concept designs for new experiments. 

The procedure of adjusting the parameters of the simulation to measured
data is called tuning. Various contributions for the tuning of MCEGs have been made 
\citep{Barate:1996fi,Hamacher:1995df,Abreu:1996na,Skands:2009zm,Buckley:2009bj,Skands:2014pea,Khachatryan:2015pea,Ilten:2016csi,Buckley:2018wdv},
 and the importance of these studies can be deduced from the recognition received.
More recently, new techniques have been presented that can improve the performance of tuning \citep{Bellm:2016voq,Mrenna:2016sih,Bothmann:2016nao,Bothmann:2018trh,Andreassen:2019nnm}.
To be able to perform the comparison of simulation and data, the data needs to
be collected and it needs to be possible to analyse the simulations similar to
the experimental setup. Here, the \texttt{hepdata} project 
\citep{Maguire:2017ypu} and analysis programs like 
\rivet{} \citep{Buckley:2010ar} are of great importance to
the high energy physics community. Once the data and the possibility to analyse
is given, the 'art' of tuning is to choose the 'right' data, possibly enhance the
importance of some data sets over others, and to modify the parameters of the
simulation such to reduce the difference of data and simulation.  A prominent
tool to allow the experienced physicist to perform the tuning is the 
\professor{} \citep{Buckley:2009bj}
package that allows performing most of the procedure automatically. 

The complexity of the MCEG tuning depends on the dimension of the parameter space used as an
input to the event generation. Further, the measured observables are in general 
functions of many of the parameters used in the simulation.  In this
contribution, we address the problems of high dimensional parameter
determination. We propose a method to choose subsets of parameters to reduce
the complexity.
We further aim to automatize the tuning process, to be able to
retune with minimal effort once improvement is made to the MCEG in use. 
We call this automation of the tuning process and the algorithm to perform it the
\texttt{Autotunes} method\footnote{An implementation of the method will be 
made available on: 
\href{https://gitlab.com/Autotunes}{https://gitlab.com/Autotunes} }. 
As possible real life scenarios we then tune the \herwig{} and \pythia{} 
models and also a hybrid form, namely the \herwig{} showers with the \pythia{}'s 
Lund String model \citep{Andersson:1983ia,Sjostrand:1984ic}.

We structure the paper as follows: 
In \Cref{sec:theproblem} we define
the problem and questions that we want to solve and answer.
We then describe \professor{} and its the capabilities and restrictions.  In 
\Cref{sec:algorithm} we explicitly define the algorithm and point out how the
methods used will act mathematically. In \Cref{sec:testing} we show how the
algorithm was tested.  Results of tuning the event generators \herwig{} and \pythia{} 
are presented in \Cref{sec:results}.  We conclude in 
\Cref{sec:conclusion} and specify the possible next steps.

\section{Current State}
\label{sec:theproblem}

Monte Carlo event generators provide theoretical predictions based on different
physics aspects. Some of these, like the generation of the hard process, are
derived from 'first principles' and include just a few parameters like the
coupling strength. The implementation of parton showers involves a number of
physics choices, like the ordering variable, which can affect the predictions.
Due to the breakdown of the perturbative description of QCD at low energy
scales, a transition to the non-perturbative regime has to be implemented, and
some more parameters are involved. Other aspects, like the hadronisation or the
description of multiple parton interactions in hadronic collisions, are based on
physical models that cannot be derived from first principles, and rely on more
parameters that have to be chosen to best describe high energy collisions. 

Improving the choice of parameters --  commonly referred to as tuning -- is required
to produce the most reliable theory predictions. The \rivet{}
toolkit allows comparing Monte Carlo event generator output to data from a
variety of physics analyses. Based on this input, different tuning
approaches can be followed. A most elaborate approach is the tuning 'by hand'.
It requires a thorough understanding of the physical processes involved in the
generation of events and the identification of suitable observables to adjust
every single parameter. 
A detailed example of such a manual approach is given by the Monash tune
 \citep{Skands:2014pea}, the current default tune of the event generator 
 \pythia{} \citep{Sjostrand:2006za,Sjostrand:2014zea}. 
 However, in order to simplify and systematize tuning efforts, a more 
 automated approach is desirable. The \professor{} \citep{Buckley:2009bj} tuning 
 tool was developed for this purpose. 
This allows to tune multiple parameters simultaneously.

\subsection{\professor{}: Capabilities and Restrictions}
\label{sec:CapabilitiesandRestrictions}

The \professor{} method of systematic generator tuning is described in detail in
\citep{Buckley:2009bj}. The basic idea is to define a goodness of fit function
between data generated with a Monte Carlo event generator and reference data that is
provided by experimental measurements through \rivet{}. This function is then
 minimized. Due to the high computational cost of generating events, a direct
evaluation of the generator response in the goodness of fit function should be avoided. 
This is done by using a parametrization function, usually a
polynomial, which is fitted to the generator response to give an interpolation
which allows for efficient minimization. The following $\chi^2$ measure is used
as a goodness of fit function between each bin $b$ of observables $\mathcal{O}$
as predicted by the Monte Carlo generator $f^{(b)}$, depending on the chosen
parameter vector $\vec p$ and as given by the reference data $\mathcal{R}_b$.
In the following, each bin in each histogram is called an observable, with prediction
$f^{(\mathcal{O})}$ and reference data value $\mathcal{R}_\mathcal{O}$:
\begin{equation}
  \chi^2 (\vec p) = \sum_\mathcal{O} w_\mathcal{O} \frac{(f^{(\mathcal{O})}(\vec p) - \mathcal{R}_\mathcal{O})^2}{\Delta_\mathcal{O}^2} \, .
\end{equation}
The uncertainty of the reference observable is denoted by $\Delta_\mathcal{O}$.
Furthermore, a weight  $w_\mathcal{O}$ is introduced for every observable.
These weights can be chosen arbitrarily to bias the influence of each
observable in the tuning process.

The approach of the \professor{} method allows to tune up to about ten
parameters simultaneously, and drastically reduces the time needed to perform a
tune.  However, further effort is needed to
overcome some of the restrictions that remain:
\begin{itemize}
  \item The polynomial approximation of the generator response is well suited for up to about ten parameters. Further simultaneous tuning requires  many parameter points as input for the polynomial fit, typically exceeding the available computing resources. This is often circumvented by identifying a subset of correlated parameters\footnote{Here and in the following we use the term correlated parameters in the sense to influence same observables. We do not discriminate between correlation or anti-correlation.} that should be tuned simultaneously.
  \item The assignment of weights requires the identification of relevant observables for the set of parameters. Different choices and methods can possibly bias the tuning result.
  \item Correlations in the data need to be identified in order to reduce the weight of equivalent data in the tune, and thus avoid bias by over-represented data.
  \item The polynomial approach is reasonable in sufficiently small intervals in the parameters, but might fail if the initial ranges for the sampled parameters are chosen too large.
\end{itemize}

\subsection{Suggested Improvements}

In the \texttt{Autotunes} approach we aim to address some of the issues mentioned above. For
high-dimensional problems, we suggest a generic way to identify correlated
high-impact parameters that need to be tuned simultaneously, and divide the
problem into suitable subsets. Instead of setting weights for every observable
by hand, we propose an automatic method that sets a high weight on highly
influential observables for every sub-tune, reducing the bias by observables
that are better optimized by parameters in another sub-tune. This procedure
makes the tuning process more easily reproducible. 

As a further improvement, we implement an automated iteration of the tuning
process, that takes refined ranges from the preceding tune as a starting point.
By a stepwise reduction of the parameter ranges, we improve the stability and
reliability of our first order approximation of parameter impact, and the
polynomial interpolation implemented in \professor{}.

\section{The Algorithm}
\label{sec:algorithm}

In this section we formulate the algorithm proposed to improve the tuning of the high
dimensional parameter space. 
We propose to organize the algorithm as:
\begin{enumerate}[label=\Alph*.]
\item Reduce the dimensionality of the problem by splitting the 
parameters into subsets, defining sub-spaces and sub-tunes. 
Here the algorithm should cluster parameters that are correlated.
\item Assign weights to observables, such that the current sub-tune
predominantly acts to reduce the weighted $\chi^2$ calculation for the 
corresponding sub-space.
\item Run \professor{}  on the sub-tunes.
\item Automatically find new parameter ranges for an iterative tuning.
\end{enumerate}

\subsection{Reduce the Dimensionality (Chunking)}
\label{sec:chunks}

\begin{figure}
\includegraphics[width=0.45\textwidth]{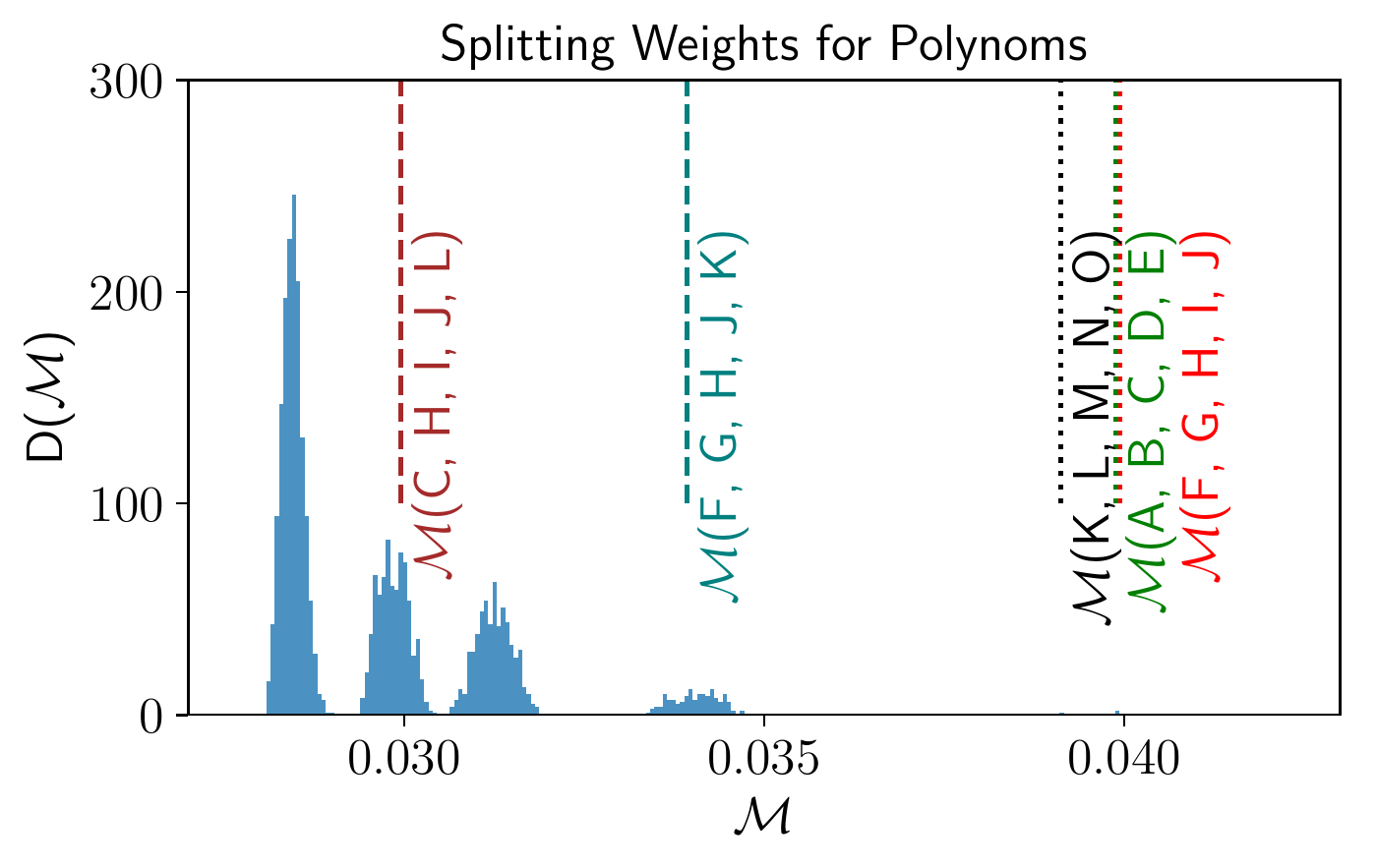}
\includegraphics[width=0.45\textwidth]{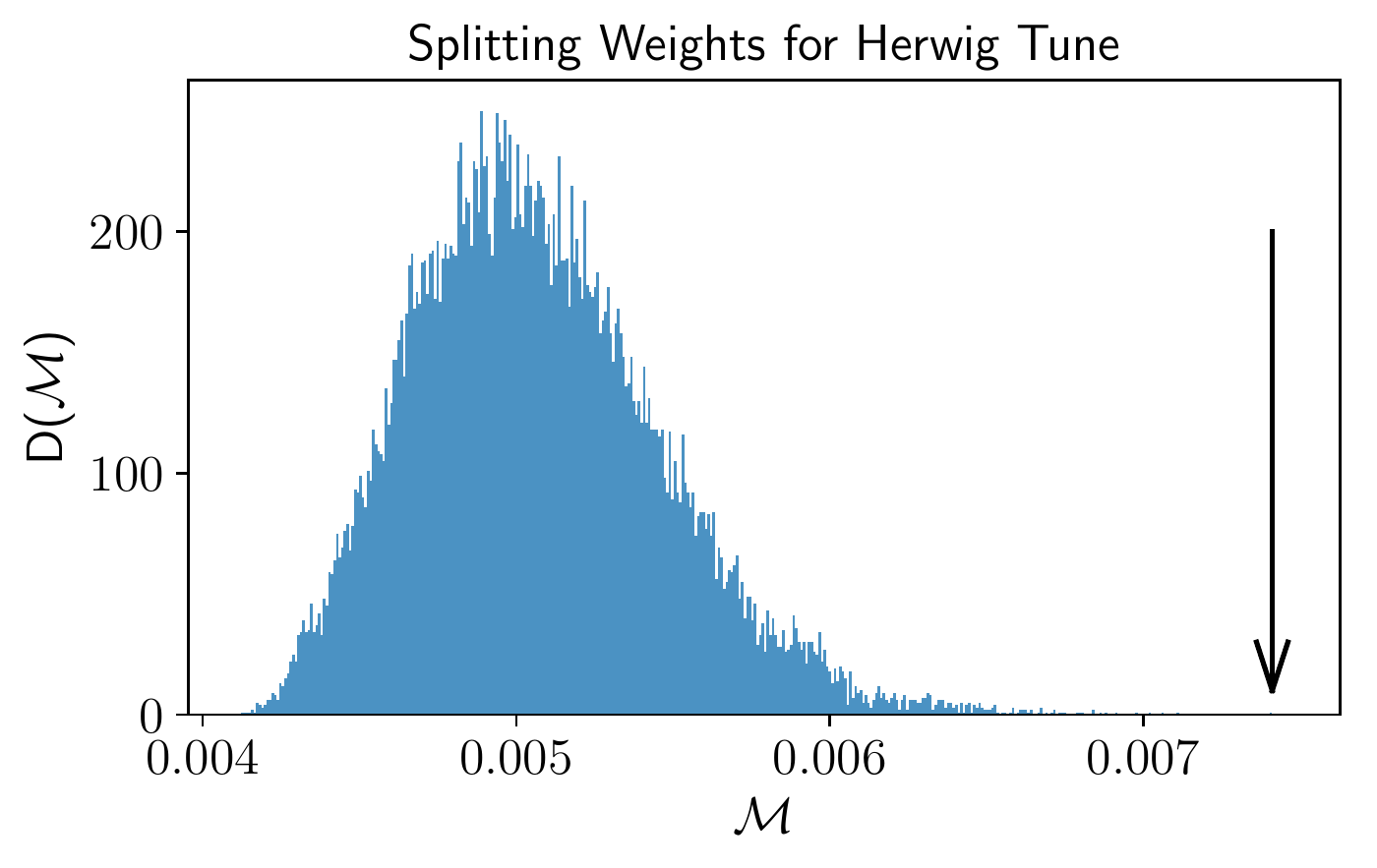}
\caption{Left: Measure density for all combination vectors according to \Cref{eq:measure} for ideal conditions of polynomial function, see \Cref{sec:test-poly}. Right: Measure density real MCEG tuning. Although the distribution is less resolving, a structure is visible. \label{fig:random-phys}
}
\end{figure}

The goal of this step is to split up a high dimensional space ($N$ dimensional) into subspaces ($n$ dimensional
\footnote{Here, the dimension $n$ is chosen such that the \professor{} package can easily manage the given subspace.}),
such that the clustered parameters are correlated on the observable level.  
To achieve this we have to define a quantity $\mathcal{M}$ that can be maximized or minimized
to allow the algorithmic treatment.  The parameter space we work with is a
hyper-rectangle. The observable definitions usually allow 
to access one dimensional projections. Here, the 'projection' is the
model (implemented in an event generator) at hand. 

Two issues directly come to
mind:
 First, we explicitly describe the parameter space $\vec{x} \in
[\vec{x}_{\min},\vec{x}_{\max}]$ as a hyper-rectangle rather than a hyper-cube.
Some of the parameters could have been measured externally, others
are pure model specific. A measure, which allows comparisons
between the parameters, needs to be corrected for the initial ranges ($[\vec{x}_{\min},\vec{x}_{\max}]$)
defined by the input.  To overcome this first problem, we first define
$\bar{x_i} \in [0,1]$ as the vector normalized to the input range and will 
describe below how a rescaling is performed to regain the information lost 
by this normalisation and relate it to the variations on the observables.   

The second
issue is the generic observable definition.  Some of the observable bins
 are parts of normalized distributions, or even related to other
histograms (as  is the case for e.g. centrality definitions in heavy ion
collisions \citep{Aad:2015zza}).  In other words, the height $y_O$ of observables again does not
define a good measure to define a generic quantity to minimize.  In order to
overcome the second problem, we test the observable space with $N_{search}$
random points in the parameter space projected with the model to the
observables. The spread for each observable is used to normalize the
values to $\bar{y}_O \in [0,1]$.  
Note that an influential parameter can be shadowed by a less important 
parameter if the latter has a too large initial range.
After the normalizations
$\bar{x}_i$ and $\bar{y}_O$ are performed, we use the $N_{search}$-projections
to perform linear regression fits for each parameter, and for each observable bin.
Due to the normalization of the $y_O$-range, the slope is influenced not only by
the parameter itself, but also by the spread produced by the other parameters.  The
reduction of the slope includes a correlation of parameters to other
parameters on the observable level.  
We use the absolute value\footnote{The
later normalization of $\vec{\mathcal{S}}_i$ but also the later definition of
$\mathcal{M}$ requires the absolute value.} of the slope to define an averaged
gradient or slope-vector $\vec{\mathcal{S}}_i$. 
 The sum $\vec{\mathcal{S}_N}=\sum_i
\vec{\mathcal{S}}_i$ has in general unequal entries, one for each  parameter in the tune.
This indicates that the input ranges $[\vec{x}_{\min},\vec{x}_{\max}]$ are of unequal influence on the observables.  
To correct for this
choice and to improve the clustering of parameters with higher correlation, we
normalize each $\vec{\mathcal{S}}_i$ element-wise with $\vec{\mathcal{S}}_N$ to
create $\vec{\mathcal{N}}_i$,
\begin{equation}
\mathcal{N}^j_i =\frac{\mathcal{S}^j_i}{\mathcal{S}^j_N}\;.
\end{equation}
 In bin $i$ the component to a parameter of the
new vector $\vec{\mathcal{N}}_i$ is reduced if other observables are sensitive
to  the same parameter.  The direction of $\vec{\mathcal{N}}_i$ indicates the
correlation of parameters . We can now use $\vec{\mathcal{N}}_i$ to chunk the
dimensionality of the problem.  Therefore, we calculate the projection for each
of the $\vec{\mathcal{N}}_i$ on all possible $n$ dimensionnal sub-spaces. This is done by
multiplication with  combination vectors $\vec{\mathcal{J}}$. 
Here, $\vec{\mathcal{J}}$ is defined as one of all possible $N$-dimensional vectors
with $N-n$ zero entries and $n$ unit entries, where $n$ is again the dimension of the desired 
sub-space, e.g. $\vec{\mathcal{J}}=(1,0,0, ... ,1,0,1)$. The sub-space then defines a sub-tune.
The sum over all projections, 
\begin{equation}
\sum_i (\vec{\mathcal{N}}_i \cdot\vec{\mathcal{J}})^k
\end{equation}
can serve as a good measure to be maximized. However, due to
the normalization of $\vec{\mathcal{N}}_i$ the sum is equal $\vec{\mathcal{J}}$  
for $k=1$. For the quantity
$\mathcal{M}$  mentioned at the start of the section we use $k=2$ giving,
\begin{equation}
\mathcal{M}(\vec{\mathcal{J}})=\sum_i  (\vec{\mathcal{N}}_i\cdot \vec{\mathcal{J}})^2\label{eq:measure}
\end{equation}
 in order to define the sub-tunes.  The maximal
$\mathcal{M}(\vec{\mathcal{J}}_{\mathrm{Step1}})$ defines the first of the sub-tunes
(Step1). For other steps, we require no overlap between the sub-spaces. This
we enforce by requiring a vanishing scalar product
$\vec{\mathcal{J}}_{\mathrm{StepN}}\cdot\vec{\mathcal{J}}_{\mathrm{StepM}}$.  
It is now possible to perform the tuning in the same order as the maximal measures of 
\Cref{eq:measure} are found. This would first fix parameters that can modify the 
description of fewer observables, and then continue to vary parameters that are 
globally important. 
In order to first constrain globally important parameters, and then fix  
specialized parameters, we invert the order of found sub-tunes.
We thus have split the dimensionality of the problem, and will ensure, in the following, that observables
used in the various sub-tunes are described by the set of influential
parameters.

\subsection{Assign Weights (Improved Importance)}
\label{sec:importance}

In the last paragraph, we described how we split up the dimensionality of the
full parameter set to allow us to tune subsets, such that parameters with higher
correlation on the observable level are tuned simultaneously.  To increase the
importance of observables that are relevant for the sub-tune, we now try to enhance the
relative weight w.r.t. other observables.  Here, we use the same vectors
$\vec{\mathcal{N}}_i$ defined in the last paragraph.  These vectors, obtained
by linear regression, and normalized to the overall range of observable
vectors have the properties, that they point in the parameter space, and, due to
the normalization, they correlate the importance of other measured observables
to the current bin. We define the weight of the observable bins later used to
minimize the $\chi^2$ as
\begin{equation}
w_i=\frac{(\vec{\mathcal{N}}_i \vec{\mathcal{J}}_\text{Step})^2}{\sum_j N^j_i} \;\label{eq:weights},
\end{equation}
where $\vec{\mathcal{J}}_\text{Step}$ is the combinatorial vector defined in 
\Cref{sec:chunks}, corresponding to the sub-tune. 
This weight has the properties that the multiplication in the numerator 
increases the weight of the important bins for the sub-tune, while the sum over 
components of $\vec{N}_i$ in the denominator reduces the importance of bins that are equally 
or more important to other parameters. Note that the $\vec{N}_i$ itself are not normalised, 
only the sum over $i$ is normalised in each component.

\begin{figure}
\includegraphics[width=0.32\textwidth]{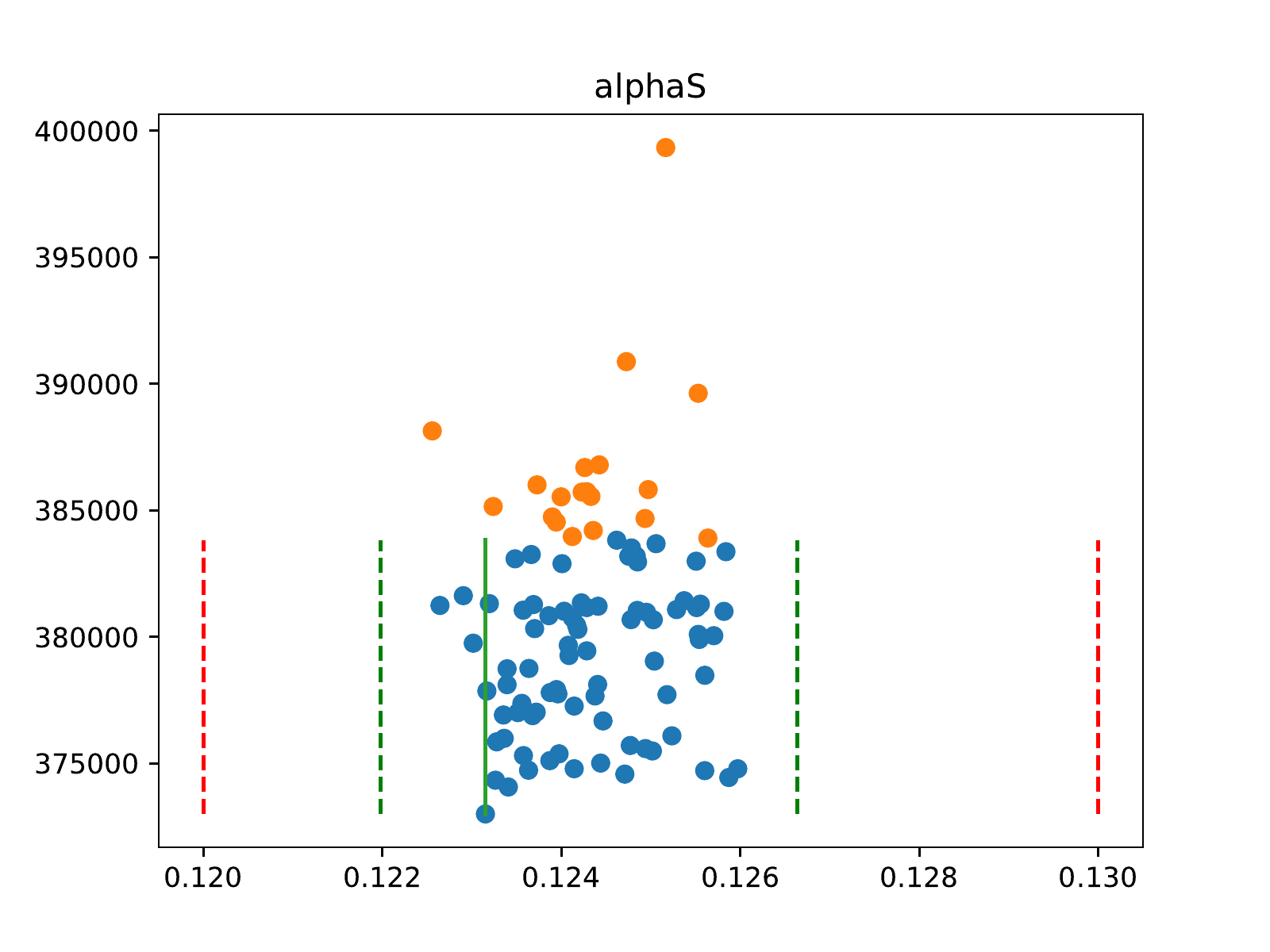}
\includegraphics[width=0.32\textwidth]{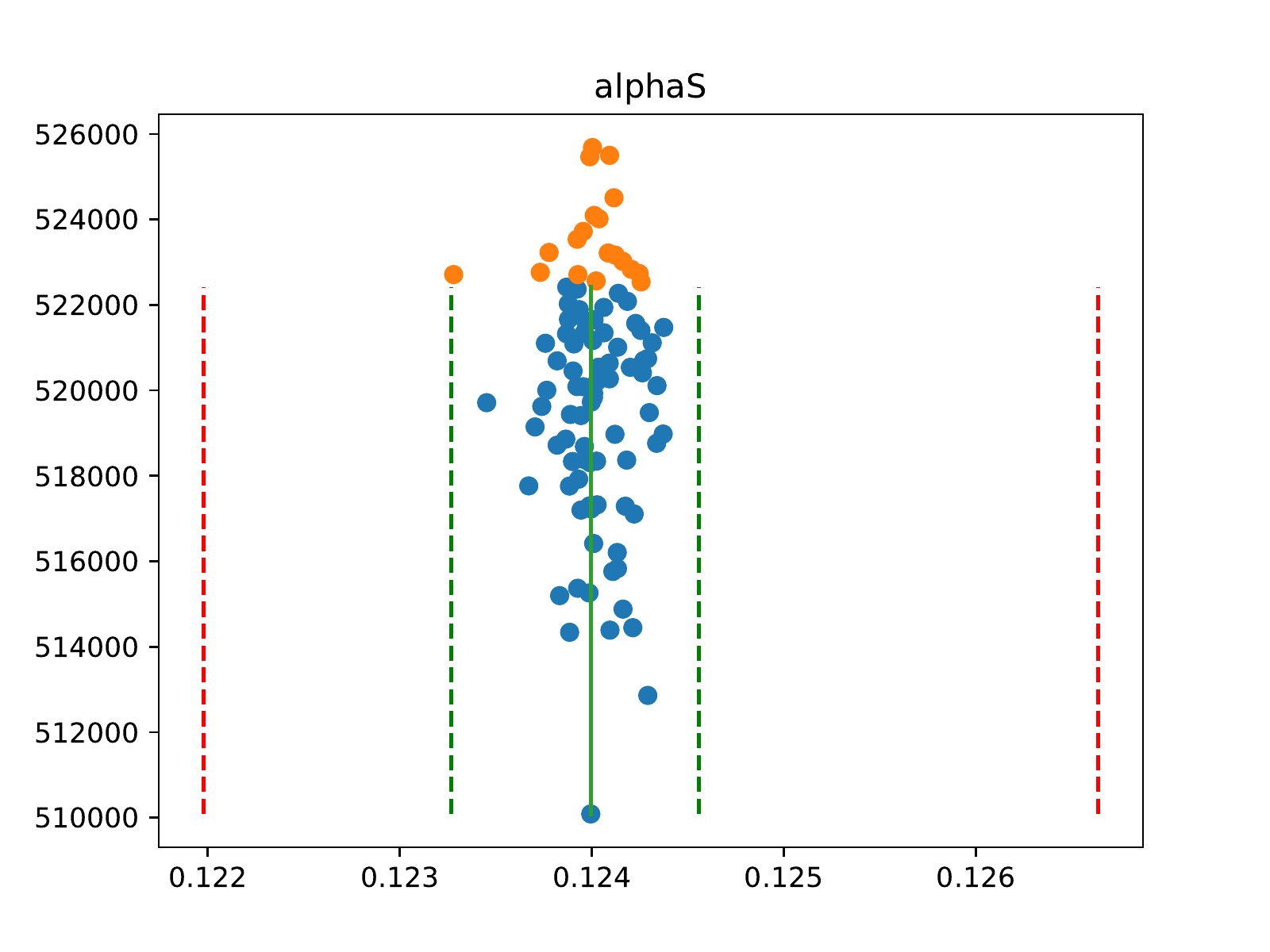}
\includegraphics[width=0.32\textwidth]{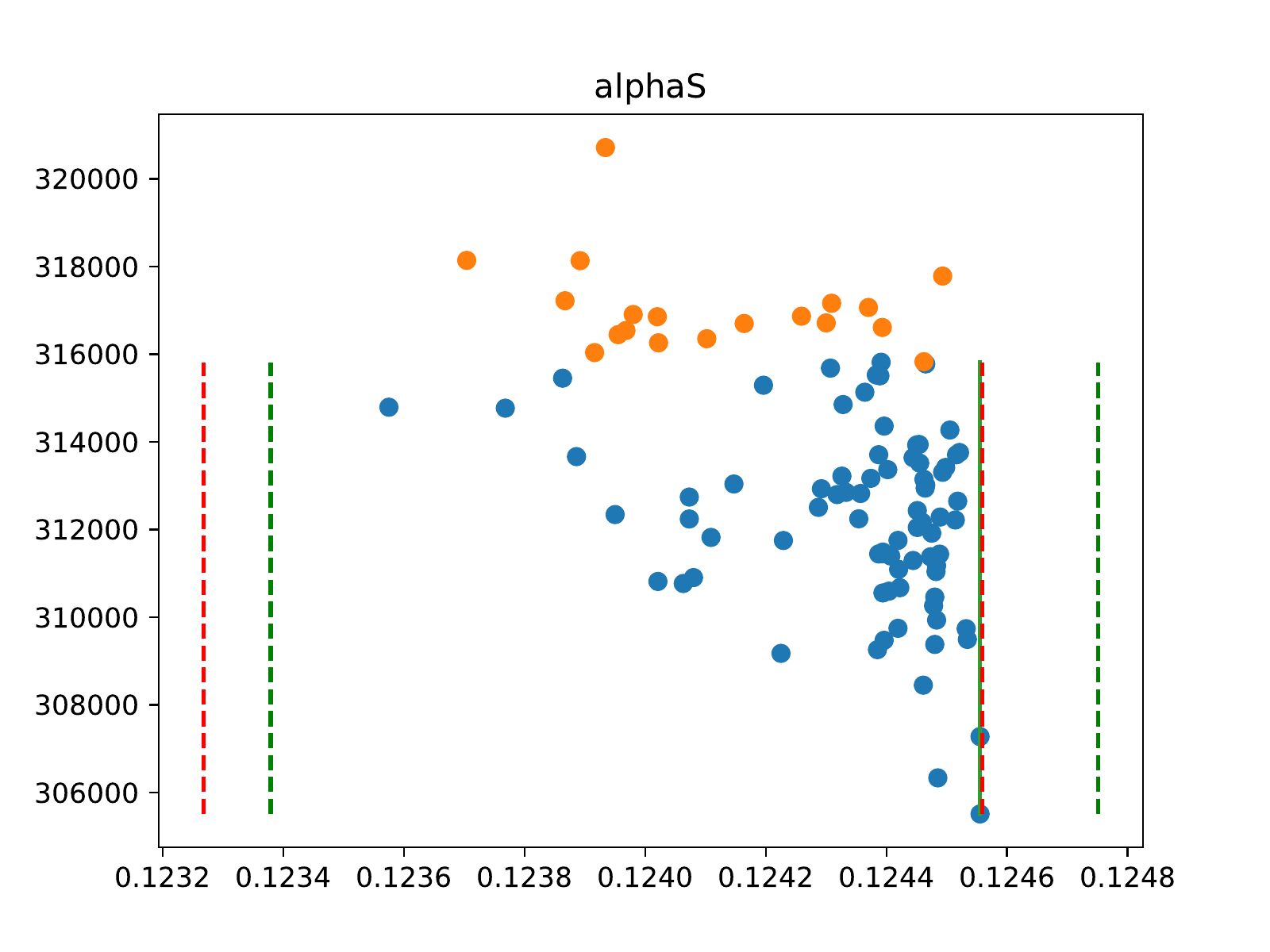}
	\caption{Example of tune results for the $\alpha_S(M_Z)$ value with three iterations from left to right. 
	The points are Goodness of fit (G.o.F.) values given by Professor for various \texttt{runcombinations}
	(see \Cref{sec:tune,sec:iterate} for details).
  While the old parameter ranges are given by dashed red lines, the 80\% of the best G.o.F. values determine the new green ranges for the next iteration. The lowest G.o.F. value defines the current best tune value that is used for next tune-steps and iterations.
   \label{fig:ranges}}
\end{figure}

\subsection{Run \professor{} (Tune-Steps)}
\label{sec:tune}

Before we start the first iteration and step, we perform a second order 
\professor{} tune as starting condition, referenced to as \texttt{BestGuessTune}. 
This is done to reduce the user interference and make use of the sampled 
points used to determine the spitting of the parameter space and the weight 
setting described in the previous sections. 

After splitting the parameter space and enhancing the weights for important
observables for the sub-tunes we use the capability of \professor{} to tune the
parameter space of each step. When a step is performed, we use the \professor{}
result of this and all previous steps to fix the parameters for the following
step.

For the individual sub-tunes, we make use of the \texttt{runcombination} method of \professor{}, to build
subsets of the randomly sampled parameter points. This produces modified polynomial interpolations 
and gives a spread in the $\chi^2$ values of the best fit values.
We choose the result associated with the best $\chi^2$ as the best tune value.
To give a measure for the stability of the tune, we choose the  \texttt{runcombinations} 
that give the best $80\%$ of the $\chi^2$ values. 
For those we  extract the corresponding parameter range, and add a $20\%$ margin on both sides. 
To elucidate the effect, an example for the tuning of the strong coupling constant $\alpha_S$ is given in \Cref{fig:ranges}. 
Here, the blue points correspond to the $80\%$ best combinations and the green dashed lines give 
the measure of stability. Diagrams like  \Cref{fig:ranges} are automatically produced 
by the program, for each parameter and tune-step. In \Cref{fig:ranges} three iterations are shown as it is described in the next section.

\subsection{Find new ranges and iterate the procedure (Iteration)}
\label{sec:iterate}

The measure of stability defined in the \Cref{sec:tune} also serves as input for the next iterations. 
Here we make use of the redefined ranges. 
An iterative tuning is important, since the first set of parameters has
been influenced by the users choices, and a next iteration can have significant 
impact on the parameter value. 
For very expensive simulations,
at least a retuning of the first step's parameter space seems desirable. The
program is setup such that one can use the output of the first full
tune as input for the next iteration.

\section{Testing and Findings}
\label{sec:testing}

Before applying the \texttt{Autotunes} framework to perform a LEP retune of \pythia{},
\herwig{}, and a combination of both in \Cref{sec:results}, we test the method
under idealized conditions. First, we tune the coefficients of a set of
polynomials. The observables used for the tune are constructed from the
polynomials for a random choice of coefficients, see \Cref{sec:test-poly}. As a second test, we tune the
\pythia{} event generator to pseudo data generated with randomized parameter
values. In both scenarios, it is desirable to recover the randomly chosen
parameter values that were used to generate the observables.

\subsection{Testing the algorithm under ideal conditions}
\label{sec:test-poly}
\begin{figure}
  \centering
  \includegraphics[width=0.45\textwidth]{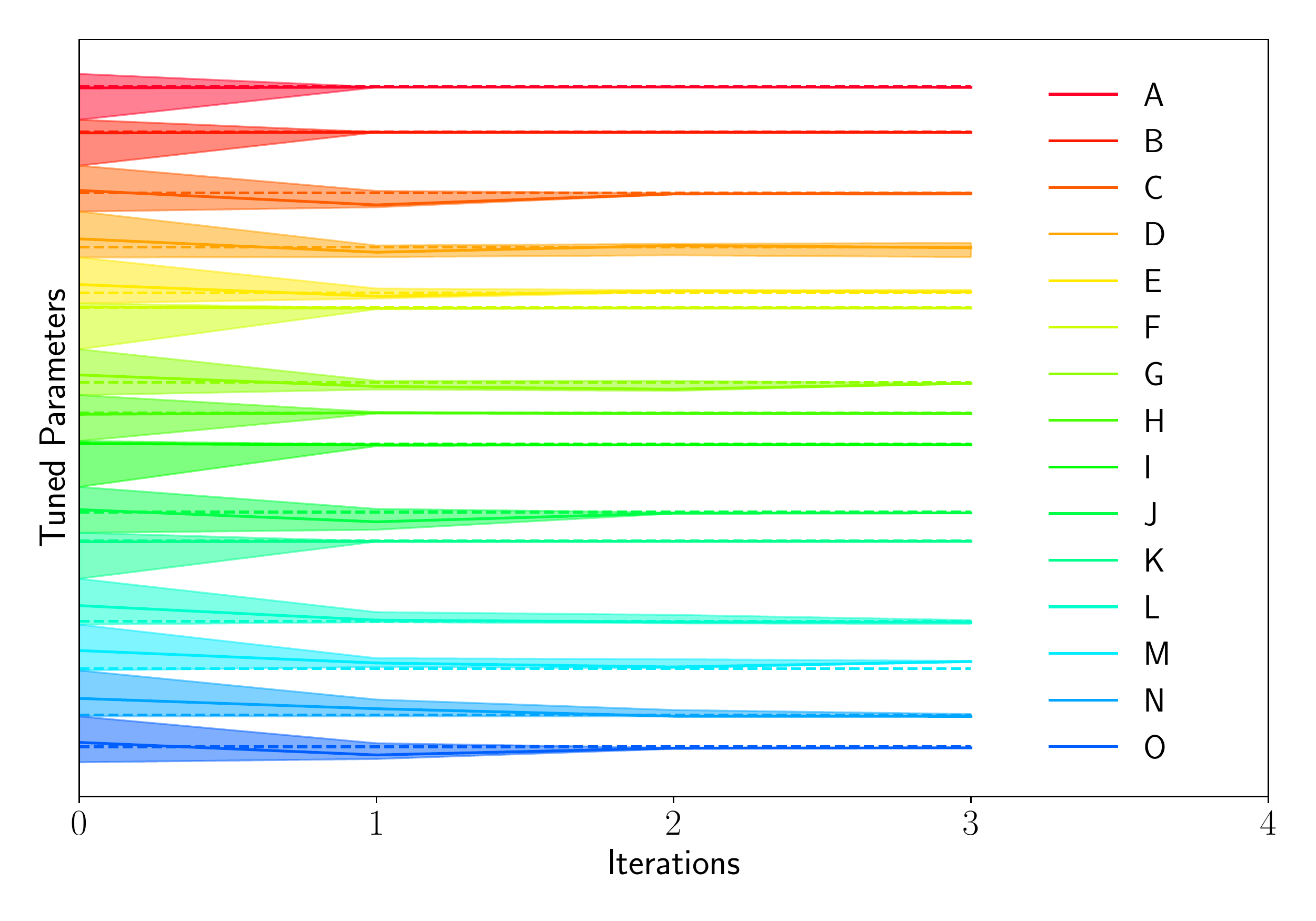}
  \caption{Iterated tuning to polynomial pseudo data using the \texttt{Autotunes} method.\label{fig:pseudo3}}
\end{figure}

To test the algorithm, we first introduce a simplified and fast generator. We define the projection,
\begin{equation}
\mathcal{O}_a=G_{0,a}+G^i_{1,a}C_a^{ir}p^r+G^{ij}_{2,a}C_a^{ir}p^rp^j+G^{ijk}_{3,a}C_a^{ir}p^rp^jp^k+G^{ijkl}_{4,a}C_a^{ir}p^rp^jp^kp^l \;\label{eq:polynom} ,
\end{equation}
with $m$-dimensional tensors $G^{\cdots}_{m,a}$, correlation
matrices\footnote{As mentioned before, we understand the correlation of parameters only on the level of influencing similar observables. 
It is therefore a simple choice to enhance subsets of parameters in the way described without off-diagonal entries in the correlation matrices.}  $C_a^{ir}$, and 
parameter points $p^i$.  Upper indices sum over the parameter dimensions. We fill $G^{\cdots}_m$ 
with random numbers and use $C_a^{ir}$ to correlate subsets of parameters. Here $C_a^{ir}$ 
is a diagonal matrix with constant entries $k>1$ if the bin $a$ should be enhanced for 
this parameter $i$ and one if not. By building ranges, we can define enhanced 
parameter sets. As an example, we use a $d=15$ dimensional parameter space, and
correlate the parameter in combinations as [A,B,C,D,E], [F,G,H,I,J] and [K,L,M,N,O].
Under these ideal conditions, we search for the correlations with the procedure described in \Cref{sec:chunks} .
In \Cref{fig:random-phys}(left), the weights for the parameter correlations are shown. The ideal combinations 
defined above create the highest weights, and would therefore be detected as correlated by the algorithm. 
In a real life MCEG tune the correlations are much less pronounced. 
In the right panel of \Cref{fig:random-phys}, we show the weight distribution for the example of the \herwig{} tune described in \Cref{sec:HwCluster}.
Once the correlated combinations are found, the algorithm continues with the procedure described in \Cref{sec:tune,sec:importance}.
As the result of each full tune serving as input to a next iteration, it is possible to 
visualize the outcome as a function of tune iterations. 
\Cref{fig:pseudo3} shows this visualisation as produced by the program.
Each parameter (A-O) is normalised to the initial range, and plotted with an offset. 
In this example, it is possible to show the input values of the pseudo data with dashed lines.
This is not possible when tuning is performed to real data. 
As \professor{} is very well capable of finding polynomial behaviour, the parameter 
point that the method aims to find is already well constrained after the first iteration. 
However, next iterations still improve the result. This may be seen for example in the third 
and last line. 

The procedure to split the parameter space into smaller subsets, and to assign 
weights can suffer from numerical and statistical noise if we consider many observables. 
In \Cref{app:rangedep}, we discuss the range dependence and 
show that the weight distributions are fairly stable if the same parameters 
are found to be correlated. It is further possible to ask for weights, if all 
parameters should be tuned independently. From the tuning perspective this 
seems an unnecessary feature, but can help to find observables that are likely
influenced by a model parameter, e.g it is possible to identify the range
of bins where the bottom mass has influence in jet rates.

\subsection{Tuning \pythia{} to pseudo data}
\label{sec:test-pythia}

As a second test of our method, we use \pythia{} to generate pseudo data for a
random choice of 18 relevant parameter values. We then use three different
methods to tune \pythia{} to this set of pseudo data, and try to recover the true
parameters. In all methods, we divide the tuning into three sub-tunes. The first
method is a random selection of parameters out of the full set, with unit
weights on all observables. In the second method, we choose the simultaneously
tuned parameters based on physical motivation, but still use unit weights on
all observables. Finally, we use the \texttt{Autotunes} method to divide the parameters
into steps, and automatically set weights as described in \Cref{sec:algorithm}.

The choice of parameters used in the physically motivated method is given in
\Cref{tab:physpar}. The first step collects parameters that have a significant
influence on many observables, combining shower and \pythia{} string parameters.
The second step gathers additional properties of the string model \citep{Andersson:1983ia,Sjostrand:1984ic},
focusing on the flavor composition. The last step then tunes the ratio of
vector-to-pseudoscalar meson production.

\begin{table}[h]
  \centering
  \begin{tabular}{|c|c|c|}
    \hline
    Step 1 & Step 2 & Step 3 \\ \hline
    TimeShower:alphaSvalue & StringFlav:probStoUD & StringFlav:mesonUDvector \\
    TimeShower:pTmin & StringFlav:probQQtoQ & StringFlav:mesonSvector        \\
    StringZ:aLund & StringFlav:probSQtoQQ & StringFlav:mesonCvector          \\
    StringZ:bLund & StringFlav:probQQ1toQQ0 & StringFlav:mesonBvector        \\
    StringPT:Sigma & StringFlav:etaSup &                                     \\
    StringZ:aExtraSQuark & StringFlav:etaPrimeSup &                          \\
    StringZ:aExtraDiquark & StringFlav:popcornRate &                         \\
    \hline
  \end{tabular}
  \caption{Parameters chosen to be tuned simultaneously in the physics motivated tuning approach.}\label{tab:physpar}
\end{table}

\begin{figure}
  \centering
  \subfigure[Iterated \pythia{} pseudo data tune with random choice of parameter subset.]{\label{fig:random}\includegraphics[width=0.49\textwidth]{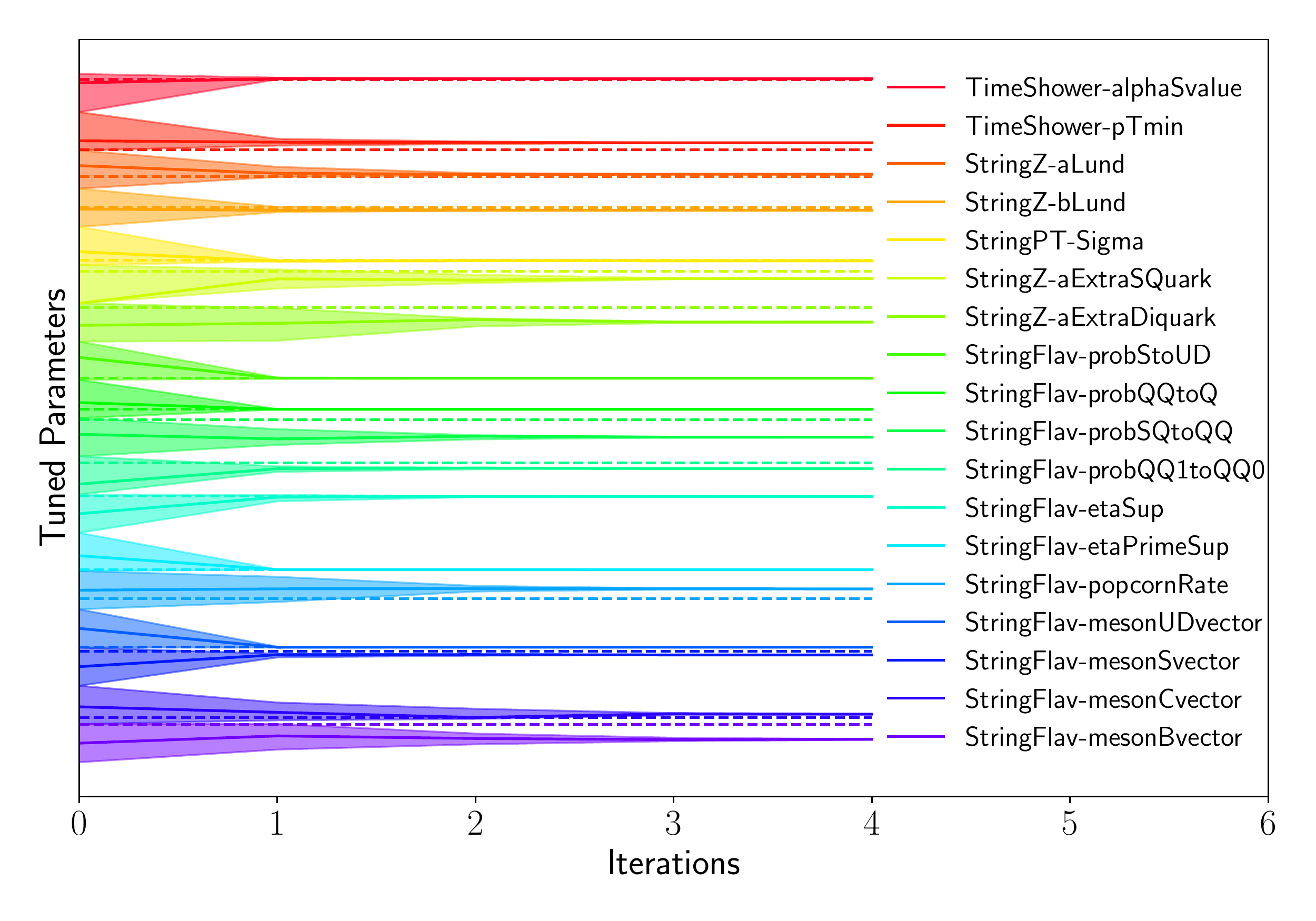}}
  \subfigure[Iterated \pythia{} pseudo data tune with physically motivated choice of parameters.]{\label{fig:phys}\includegraphics[width=0.49\textwidth]{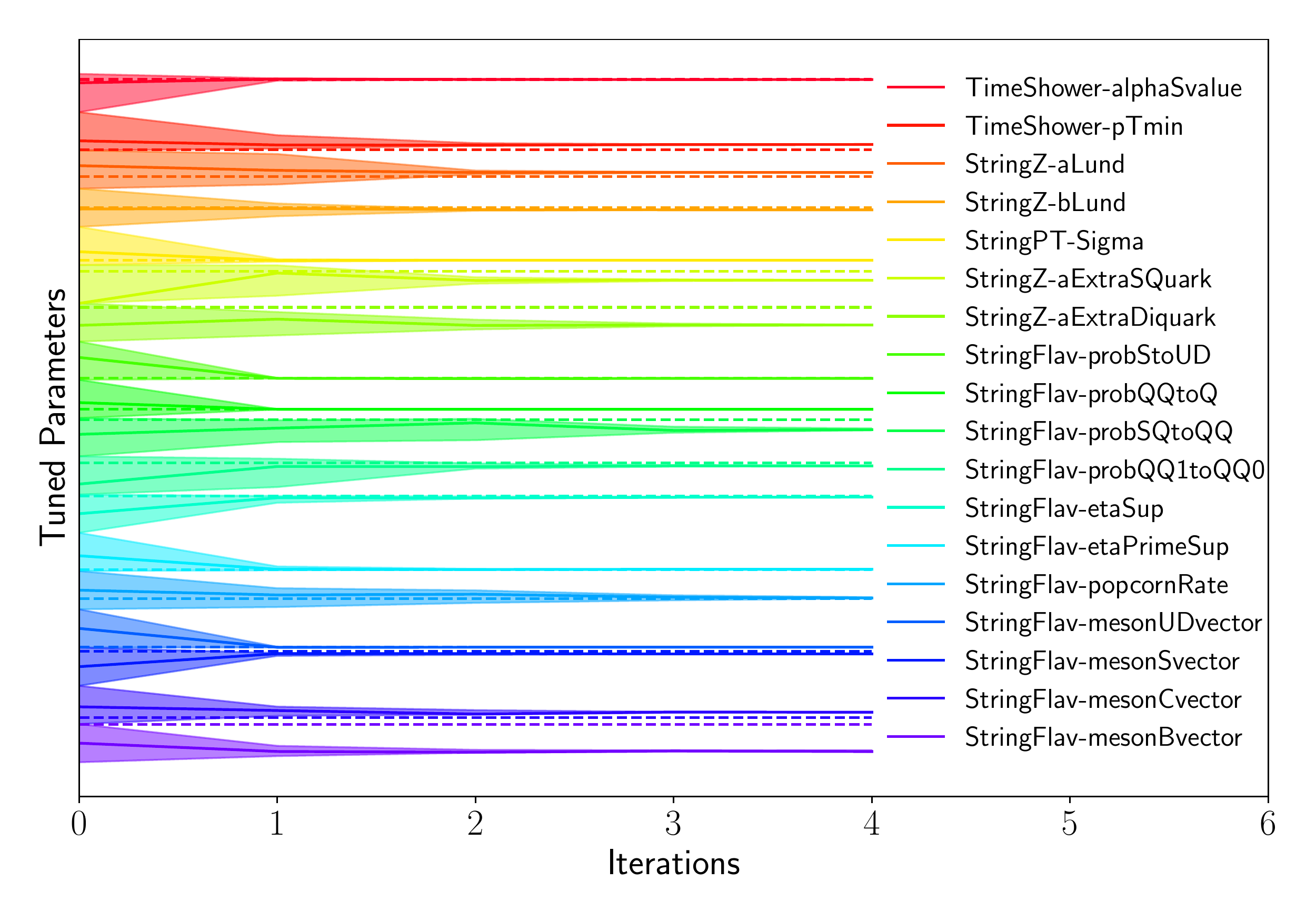}}
  \subfigure[Iterated \pythia{} pseudo data tune using the \texttt{Autotunes} method.]{\label{fig:autotunes}\includegraphics[width=0.49\textwidth]{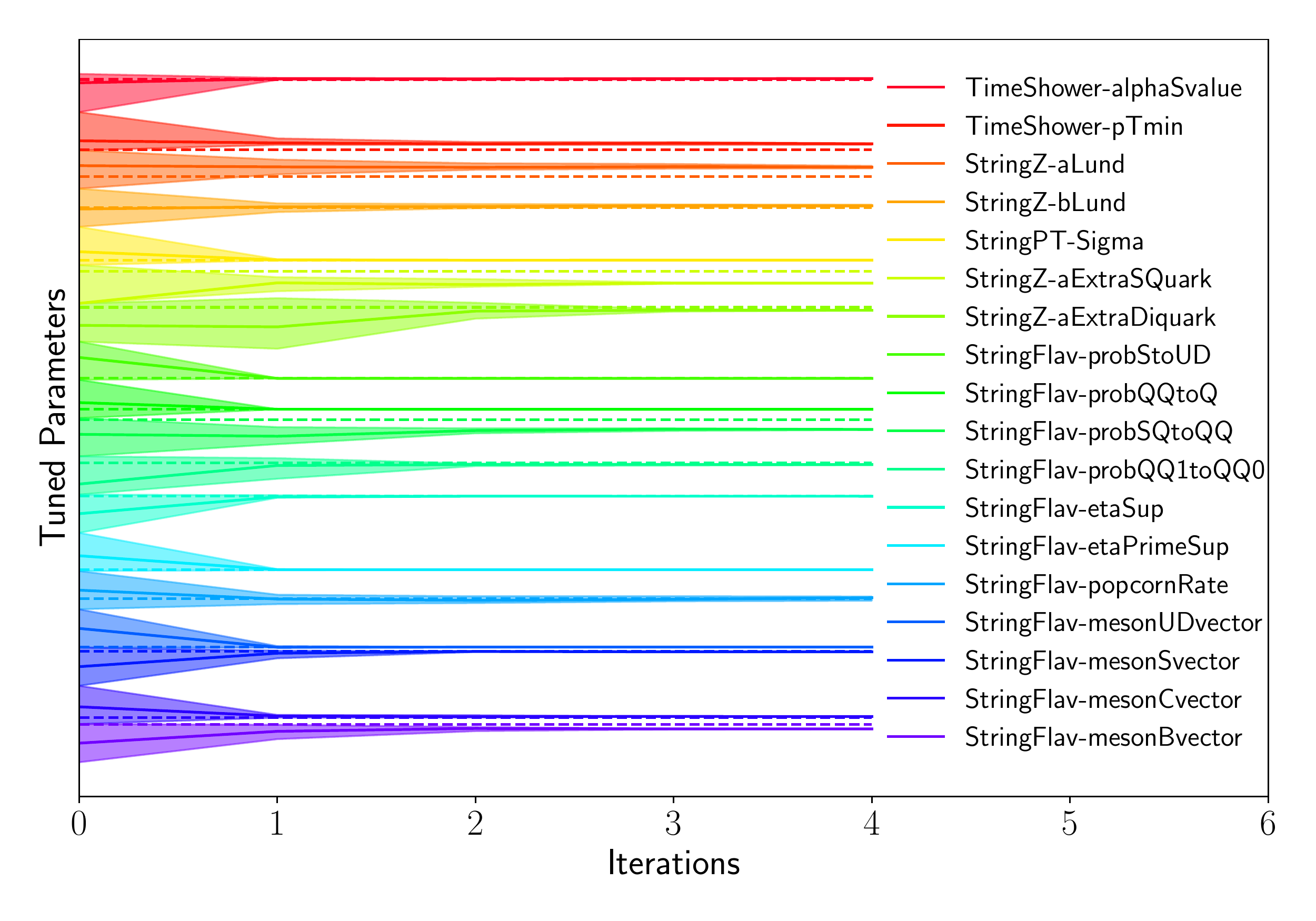}}
  \subfigure[Comparison in summed deviation from true parameters. Iterating the \texttt{Autotunes} approach leads to better agreement with initially chosen parameters.]{\label{fig:comp}\includegraphics[width=0.49\textwidth]{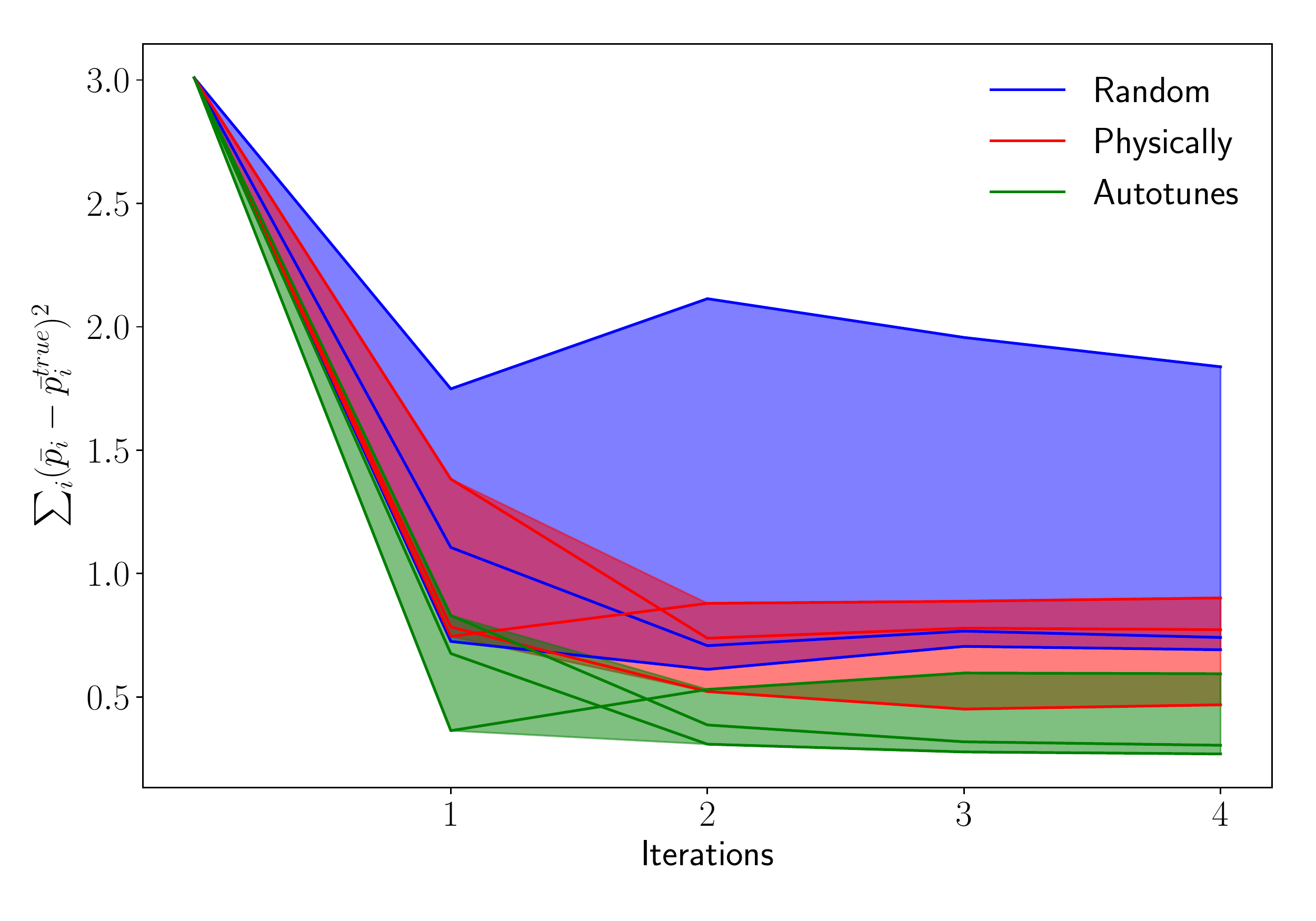}}
  \caption{Parameter development as a function of tune iterations. The dashed lines in \Cref{fig:random,fig:phys,fig:autotunes} shows the true parameter point that was used to produce the pseudo data. The uncertainty bands are given by 80\% of the best fit values in the \professor{} run-combinations and an additional 20\% margin. In  \Cref{fig:comp}  we compare the the summed deviation for three distinct tunes for the random, physically motivated and \texttt{Autotunes} method.}
\end{figure}

The results of the three tuning approaches that aim to recover the \pythia{} pseudo data
parameters are shown in \Cref{fig:random,fig:phys,fig:autotunes,fig:comp}. None
of the approaches is capable of exactly recovering all of the original
parameter values. This suggests that close-by points in parameter space are
well suited to reproduce the pseudo data observable distributions. However, the
iterated \texttt{Autotunes} method improves the agreement of the recovered parameters by
avoiding large mismatches. In the physically motivated and random approaches,
there is a certain chance that parameters are strongly constrained by
observables that also depend on other parameters. If these are not
identified and included in the same sub-tune, both parameters get constrained.
Thus, the optimal configuration is not necessarily recovered. By iteratively
identifying such sets of parameters, the \texttt{Autotunes} method avoids these
mismatches. 

\Cref{fig:comp} shows the summed, squared and normalized deviation of the
recovered to the true parameter values. Each approach is
performed three times to access the stability of the results. The random
approach uses random combinations of parameters for the tuning steps, so we see
a wide spread of results. The iterative tuning using our fixed physically
motivated parameter choice is more reliable, showing a lower spread and better
results. The \texttt{Autotunes} method leads to the best agreement with the original
parameters. More stable results in the physically motivated and the \texttt{Autotunes}
method could be achieved by using higher statistics for both the event
generation and the sampling. We see that in the physically motivated and the
\texttt{Autotunes} approach, a second tuning iteration affects the results, mostly -- but
not necessarily -- improving the parameter agreement. Further iterations have
a minor impact.

\section{Results}
\label{sec:results}

We use the \texttt{Autotunes} framework to perform five distinct tunes to LEP
observables.  We provide the list of analyses in the additional material with the arXiv
upload. To this point we do not weight the LEP observables, but make use of the
sub-tune weights described in \Cref{sec:importance}
\footnote{Additional weighting with knowledge of perturbative stability 
or known misinterpretation of experimental errors may be subject to future work.}.
The tunes make use of the default hadronisation models of the 
event generators \herwig{} and \pythia{}. We further present a new tune of the \herwig{} event 
generator interfaced to the \pythia{} string hadronisation model. The details of the simulations can be found in the following sections. 
The results are presented in \Cref{tab:pythiatune} and \Cref{tab:herwigcluster}, 
listing default values, tuning ranges of the parameters, as well as the tuning 
results using the \texttt{Autotunes} method.

\subsection{Retuning of \pythia{}}

The tune of \texttt{Pythia 8.235} is performed by using LEP data. We use \pythia{}'s standard
configuration as described in the manual, including a one-loop running of
$\alpha_\text{s}$ in the parton shower. The tuned parameters, initial ranges
and tune results are given in \Cref{tab:pythiatune} in \Cref{app:tune-results}.

The given ranges on the tune results, obtained from the variation of the optimal tune in different run combinations, 
can be interpreted as a measure of the stability of the best tune. 
A wide range suggests that different configurations give tunes of similar $\chi^2$. 
The extraction of the strong coupling $\alpha_s$ is the most stable result in the tune. 
The modification of the longitudinal lightcone fraction distribution in the string fragmentation model for strange quarks (StringZ:aExtraSQuark) is very loosely constrained, suggesting that the data that is employed in the tune is not suitable to extract this parameter.

We tune 18 parameters in three sets of six parameters each. In the \pythia{} tune, the parton shower cutoff pTmin is surprisingly loosely constrained. Checking the combinations of parameters that the \texttt{Autotunes} method chooses, we note that pTmin is found to be correlated with the string fragmentation parameters aLund and bLund in every iteration, which are also rather loosely constrained. 
This suggests that different choices for these three parameters can provide tunes of similar $\chi^2$.

\subsection{Retuning of \herwig{}}

As another real life example we tune the \herwig{} event generator to LEP data. 
Here the tune is based upon version \herwig{}.1.4 and \thepeg{} 2.1.4.
We perform two tunes -- cluster and string model -- for both showers, 
the QTilde shower \citep{Gieseke:2003rz} and the dipole shower~\citep{Platzer:2009jq}. 
For the presented tunes we do not employ 
the CMW scheme \citep{Catani:1990rr}, but keep the $\alpha_S(M_Z)$ value a free parameter.
This results in the enhanced value compared to the world average \citep{Tanabashi:2018oca}. 
 
\subsubsection{Tuning \herwig{} with cluster model}
\label{sec:HwCluster}
We retune the cluster-model with a 22 dimensional parameter space. 
Here, we require tree sub-tunes and performed four iterations. 
The results are listed in \Cref{app:tune-results}. 
Comparing the results, we note that the method  is in general able to find values 
outside of the given initial parameter ranges, see e.g. the $\alpha_S(M_Z)$ or the 
nominal $b$-mass. 
This can be caused by \professor{} interpolation outside the given bounds 
or in the determination of the new ranges for the next iteration. 
Apart from the parameters that influence the cluster fission process of heavy clusters
 involving charm quarks (ClPowCharm and PSplitCharm), the  parameters are comparable 
between the two shower models. Further in the cluster-model, the fission parameters 
are correlated. It is reasonable to assume possible local minima in the $\chi^2$ measure.

\subsubsection{Tuning \herwig{} with \pythia{} Strings}
The usual setup of the event generators are genuinely well-tuned and even
though the tests of \Cref{sec:testing} allow the conclusion that relatively
arbitrary starting points lead to similar results, ignoring the previous
knowledge completely seems undesirable.  To create a real live example and
further allow useful future studies we employed the fact that the \texttt{C++} version
of the \texttt{Ariadne} shower but also the \herwig{} event generator is based on \thepeg.
Furthermore with minor modifications, the unpublished interface between \thepeg{}
and \pythia{} (called \texttt{TheP8I}, written by L. L\"onnblad), allowed the internal use of
\pythia{}-stings with \herwig{} events. Since no tuning for this setup was attempted before
 the starting conditions needed to be chosen with less bias compared to the
other results of this section.

When we compare the values received for the \herwig{} showers to the \pythia{} shower, we note a 
comparably large value for the \pythia{} $\alpha_S$ value. In contrast, the cutoff in the
 transverse momentum in \pythia{} is rather small. 
 The reason for this contradicting 
 behaviour\footnote{Observables like the number of charged particles are both likely to be modified 
 in the same direction with an increased coupling and a decreased evolution cutoff. }  
 can be found in the order at which the two codes evaluate the running of the strong coupling. 
 While \herwig{} chooses an NLO running, \pythia{} evolves $\alpha_S$ with LO running, and
  therefore suppresses the radiation for low energies. 
Even though the shower models are rather different, the difference in the response in the best fit 
values of the parameters are moderate. 
Less constrained parameters like the popcornRate, which influences part of the baryon
 production or the additional strange quark parameter aExtraSQuark show a corresponding 
 large uncertainty. It can be concluded that the data used for tuning is hardly constraining these parameters.

\section{Conclusion and Outlook}
\label{sec:conclusion}

We presented an algorithm that allows a semi-automatic Monte Carlo Event generator 
tuning of high dimensional parameter space. 
Here, we motivated and described how the parameter space can be split into 
sub-spaces, based on the projections to and variations in the observable space. 
We then assigned increased weights when we perform the sub-tunes, such that
influential observables are highlighted. 
It is then possible to use the output of any tune step as starting conditions for 
next steps. Therefore the procedure is iterative. 
In ideal conditions, we performed tests to check that the algorithm finds correlated
parameters and showed in realistic environment that pseudo data could be reproduced 
better by the algorithm than by random or physically motivated tunes. 
As real life examples we tuned the \pythia{} and \herwig{} showers with their 
standard hadronisations models and modified the \herwig{} generator to allow 
consistent hadronisation with the \pythia{}'s Lund String model. 

The method allows to perform tuning with far less human interaction.
It also allows different models to be tuned with a similar bias. Such tunes can then 
be used to identify mismodelling, with the assurance that the origin of the difference 
in data description is less likely part of a better or worse tuning.

At the current stage we did not assign weights or uncertainties other 
than the sub-tune weights and the uncertainties given by the experimental collaborations. 
We note that the difference between higher multiplicity merged simulations to the 
pure parton shower simulations can serve as an excellent reduction weight to 
suppress observables influenced by higher order calculations. However, the investigation 
of such procedures goes beyond the scope of this paper and will be subject to future work. 
Further, we did not address the third point of the mentioned restrictions in 
\Cref{sec:CapabilitiesandRestrictions} that describes over-represented data. 
We postpone such studies, that include clustering of slope-vectors to reduce such an influence, to future work.

\appendix

\section{Range dependence}
\label{app:rangedep}

The algorithm to split the dimensions and to assign weights to sub-tunes 
is constructed such that correlations should still be found when the parameter ranges are varied. 
This is not always possible if the parameter ranges are 
strongly modified. It is possible that the slope vectors, that are evaluated by averaging 
over the full $n-1$ (other) dimensions, are modified by the newly defined initial ranges. 
It is even possible that the range of other parameters influence 
the slope as the spread modifies the normalisation. 
In order to show such behaviour (and also to illustrate the weight distributions), we 
choose three different setups for the event generator \herwig{}.
We choose $d=4$ and try to split the dimensions in half. Here we choose the parameters 
and initial ranges as, 

\begin{center}
\begin{tabular}{|l|c|c|c|}
\hline Parameter  & Setup 1 & Setup 2 & Setup 3\\
\hline$\alpha_S(M_Z)$ & 0.12 -- 0.13 & 0.124 -- 0.126 &  0.12 -- 0.13\\
$Cl_{\max}^{light}$ & 2.0 -- 3.0 &  2.4 -- 2.6  &  2.4 -- 2.6\\
$p^T_{\min}$ & 0.7 -- 0.9 &  0.7 -- 0.9 &  0.78 -- 0.82\\
$g_{CM}$ & 0.7 -- 0.9 &  0.7 -- 0.9 & 0.7 -- 0.9\\\hline
\end{tabular}
\end{center}

\begin{figure}
	\includegraphics[width=0.49\textwidth]{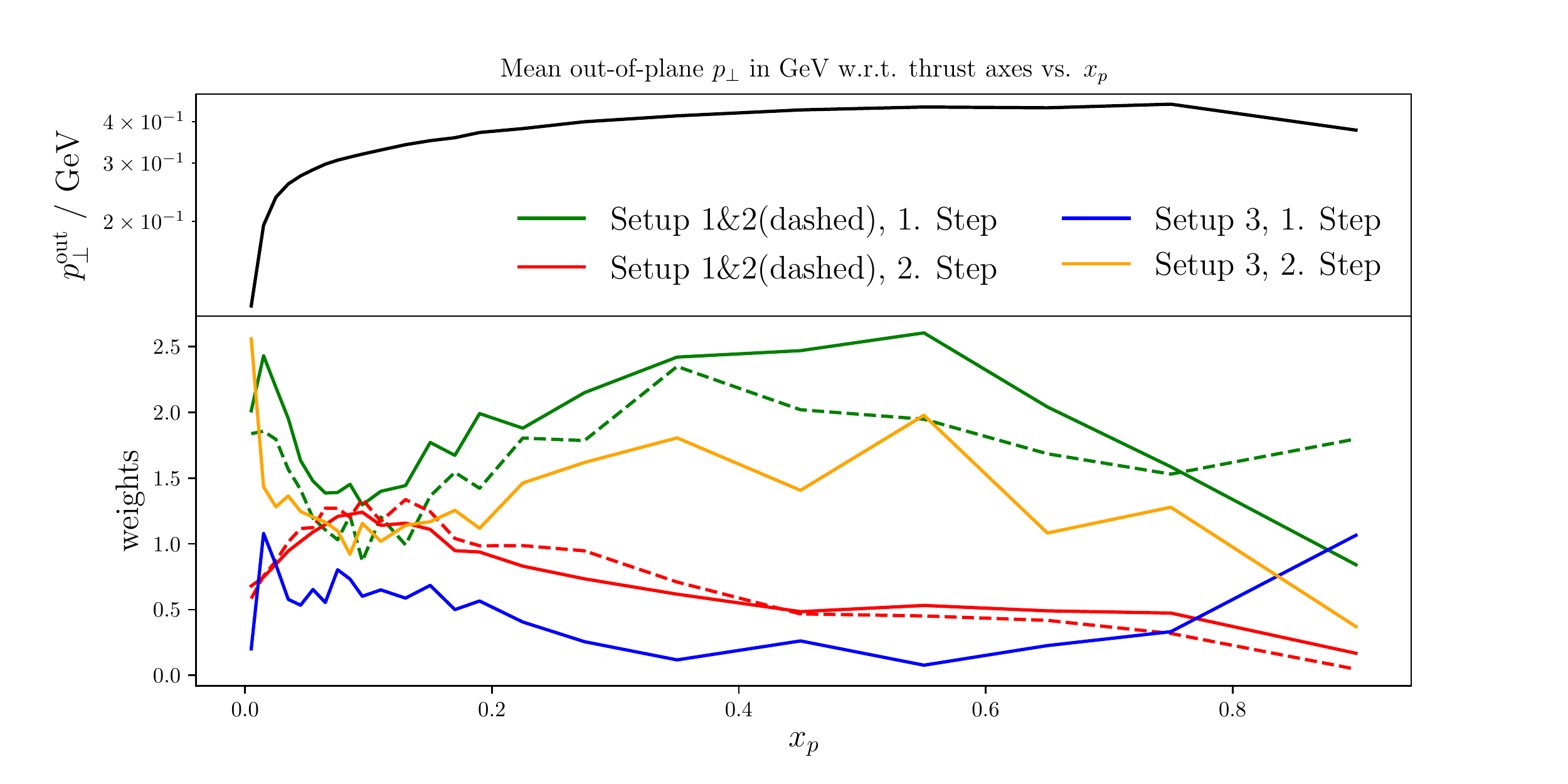}
	\includegraphics[width=0.49\textwidth]{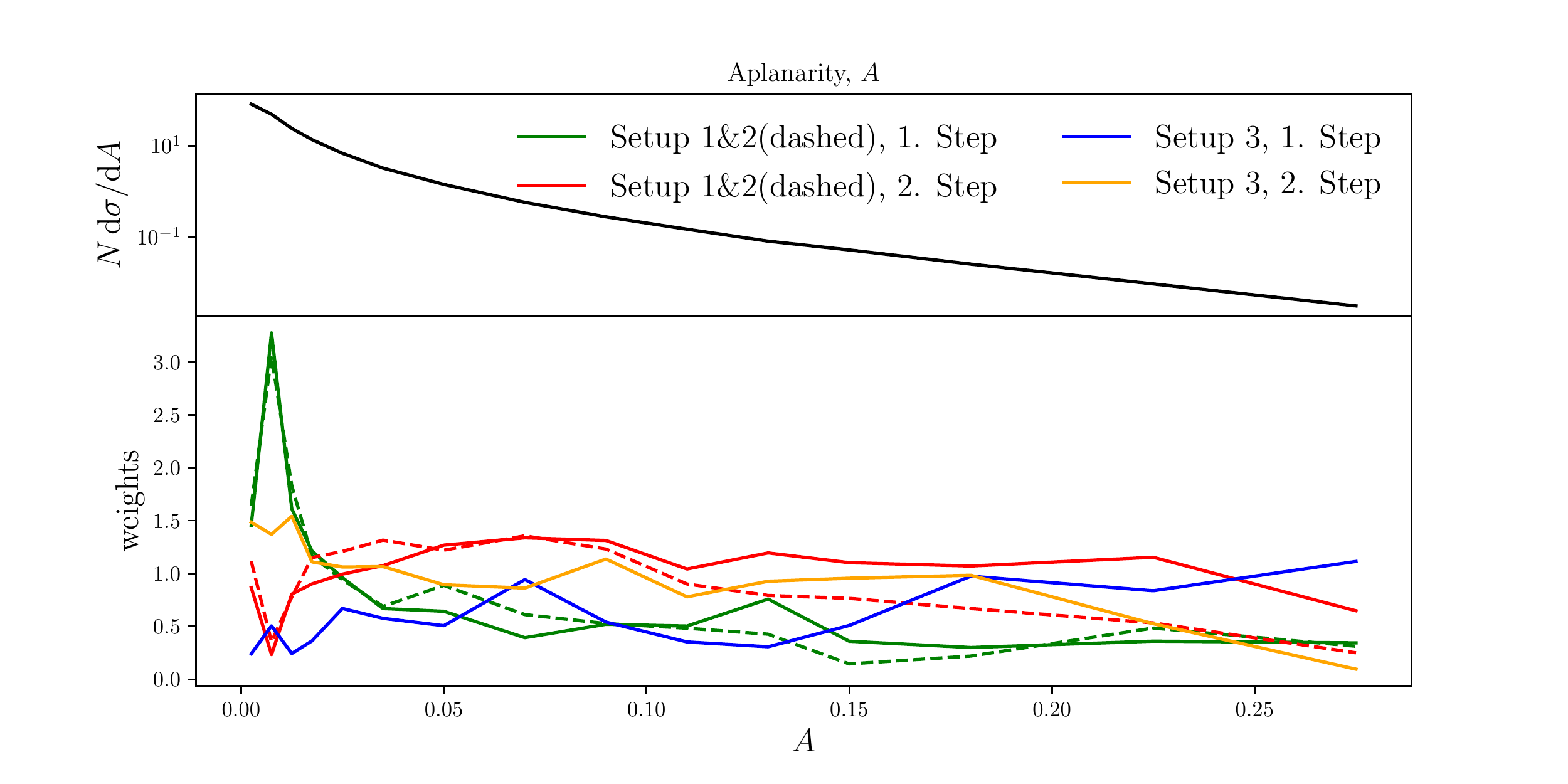}
	\includegraphics[width=0.49\textwidth]{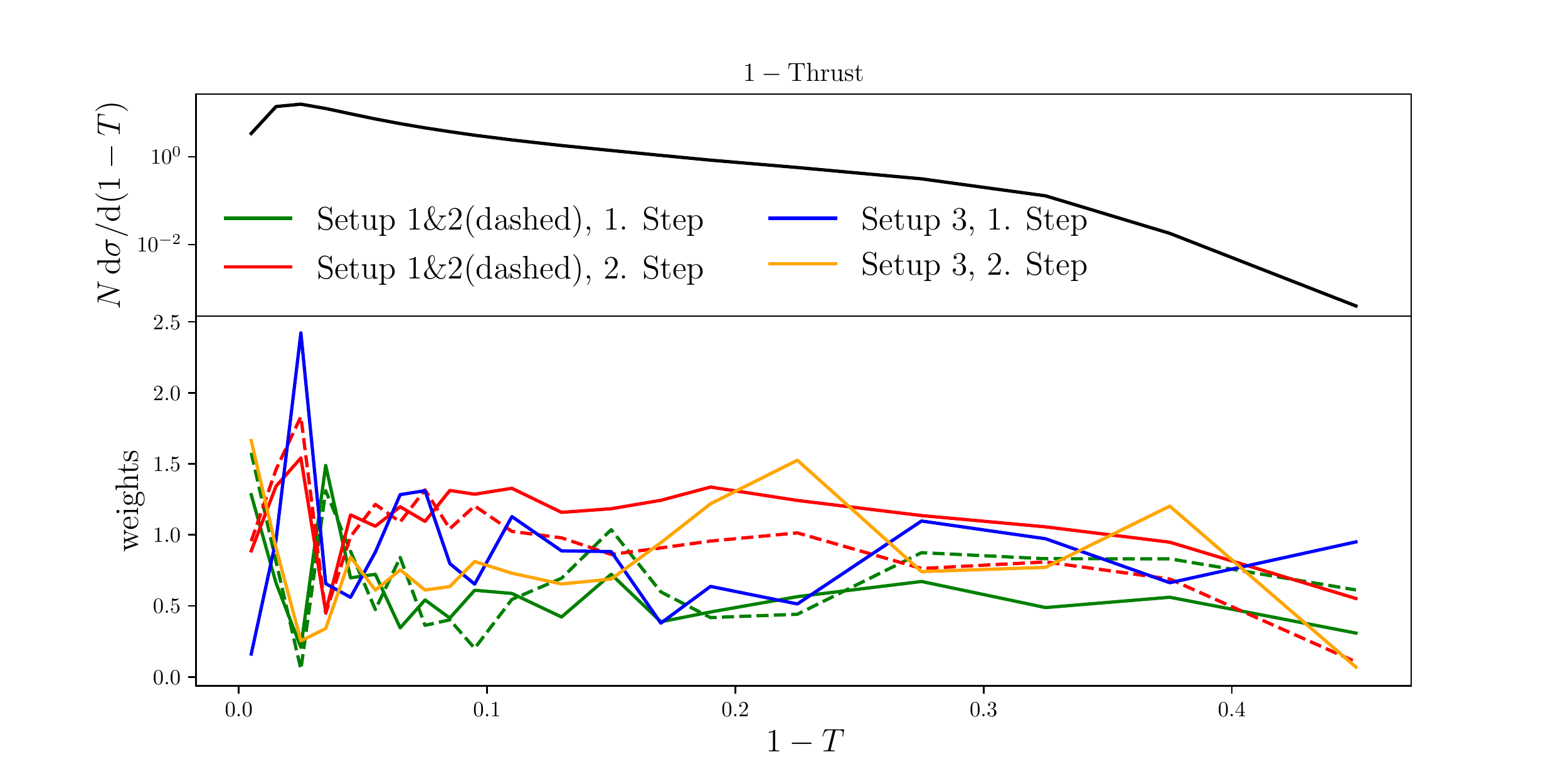}
	\includegraphics[width=0.49\textwidth]{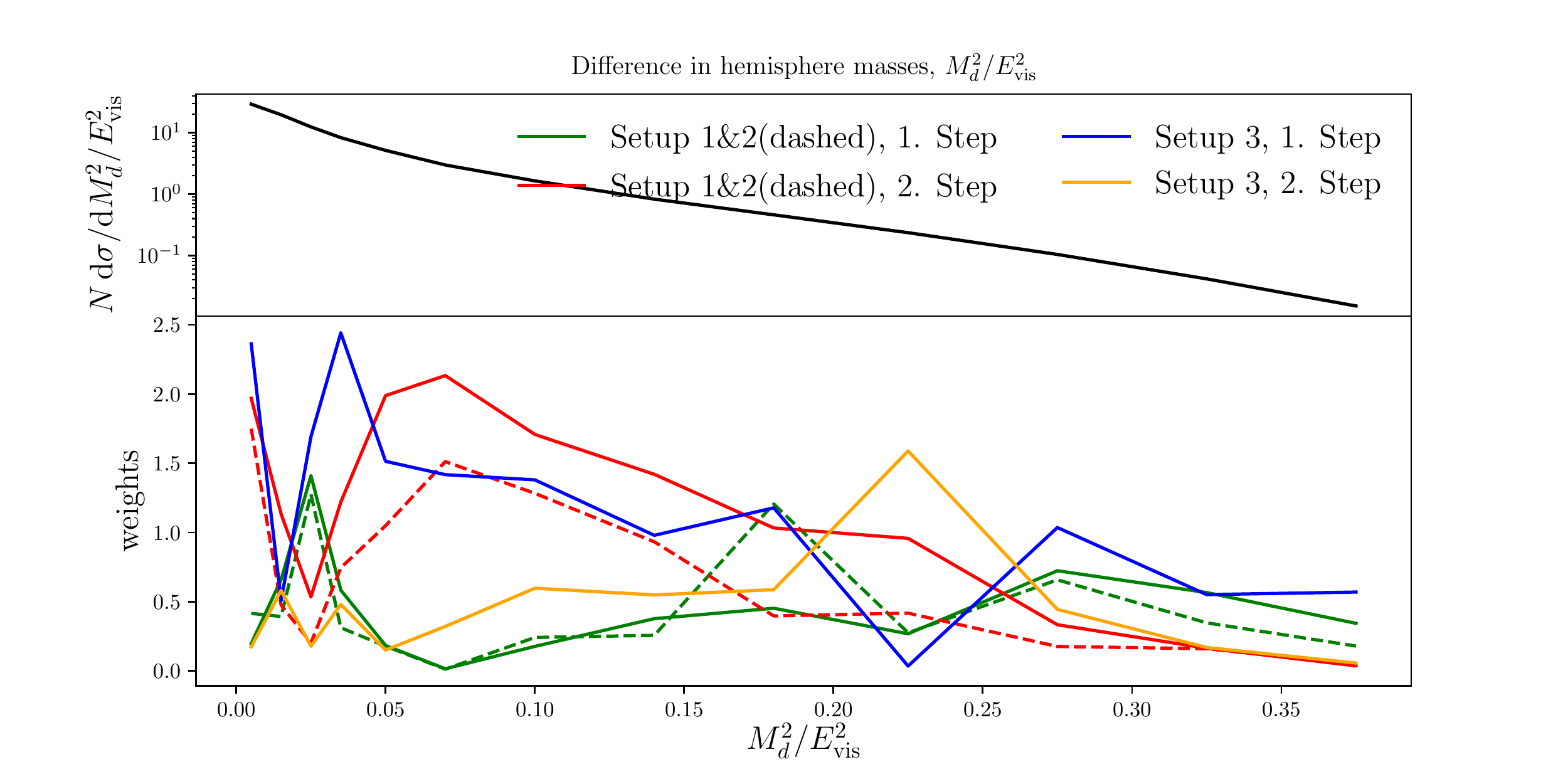}
	\caption{Weight distributions for subsets of  parameter pairs as described
	in \Cref{app:rangedep}. The upper panels show the measured data points and the lower 
	panels show the weights assigned. A clear distinction between more and less important 
	sets is visible. The dashed lines correspond to Setup 2, which gives a same grouping of 
	parameters as Setup 1.
	\label{fig:weights_from_diff_ranges}.}
\end{figure}

The result for the parameter grouping and the weight distributions are depicted in 
\Cref{fig:weights_from_diff_ranges}. 
While the algorithm to split the parameter space in  setup 1 and setup 2 
such that $Cl_{\max}^{Light}$ and  $p^T_{\min}$ should be tuned in the first step and 
then $\alpha_S(M_Z)$ and $g_{CM}$\footnote{$g_{CM}$ is the parameter for the constituent mass of the gluon.} in a second step, the modification to the initial 
ranges has the effect that the algorithm favours the pairing 
($Cl_{\max}^{Light}$ , $g_{CM}$) and ($\alpha_S$ , $p^T_{\min}$) for 
steps 1 and 2 for setup 3. 

While it is possible that by changing the initial ranges the pairing flips and 
other parameter groups are found, the fact that neighbouring bins have a similar 
behaviour supports the concept of meaningful weight distributions. 
It would be possible to correlate neighbouring bins or introduce a smoothing algorithm 
to make the weights more stable but such a modification can be introduced once issues 
with the current algorithm appear. 

In principle, it is possible to visualize for each parameter the weights of the sub-tune choice that we want. 
This choice can help to identify observables that are influential for individual 
parameters, and give insights in unexpected behaviours. Already from the weight distributions 
shown in \Cref{fig:weights_from_diff_ranges}, we can deduce that $p^T_{\min}$ is of 
great importance for the transverse momentum out-of-plane, see upper left panel. 
Further modifications of the constituent mass of the gluon $g_{CM}$  will influence the difference in the hemisphere masses, see lower right panel.

\section{Tune Results}
\label{app:tune-results}

In \Cref{tab:pythiatune} and \Cref{tab:herwigcluster} we list the results of the
 \herwig{} and \pythia{} tunes with the standard hadronisation. For \herwig{} we 
 also list a tune for the Lund string model. The results are discussed in \Cref{sec:results}.

\begin{table}[h]
  \renewcommand*{\arraystretch}{1.2}
  \begin{tabular}{|l|c|c|c|c|c|}
    \hline
    Parameter & Def. & range & \pythia{} tune & H7+$\tilde{Q}$+Str. & H7+Dip.+Str. \\ \hline 
    alphaSvalue      & $0.1365$ & $0.125 - 0.14$ & $0.13699_{-0.00057}^{+0.00019}$     & $0.1289_{-0.0004}^{+0.0011}$ & $0.13229_{-0.00015}^{+0.00083}$ \\
    pTmin            &$0.5$ & $0.4 - 0.8$ & $0.49_{-0.16}^{+0.05}$      & $0.993_{-0.004}^{+0.010}$ & $0.990_{-0.004}^{+0.020}$ \\ 
    SZ-aLund               &$0.68$ &$0.5 - 0.8$ & $0.71_{-0.24}^{+0.07}$      & $0.60_{-0.19}^{+0.13}$ & $0.84_{-0.16}^{+0.03}$ \\ 
    SZ-bLund               &$0.98$ & $0.7 - 1.3$ & $1.11_{-0.23}^{+0.11}$     & $0.78_{-0.19}^{+0.18}$ & $1.00_{-0.20}^{+0.04}$ \\
    StringPT-Sigma              &$0.335$& $0.3 - 0.4$ & $0.3011_{-0.0010}^{+0.0020}$      & $0.3008_{-0.0022}^{+0.0006}$ & $0.29876_{-0.00022}^{+0.00122}$ \\
    SZ-aExtraSQuark        &$0.0$& $0.0 - 0.5$ & $0.04_{-0.05}^{+0.60}$      & $0.08_{-0.08}^{+0.26}$ & $0.22_{-0.22}^{+0.37}$ \\
    SZ-aExtraDiquark       &$0.97$& $0.8 - 1.2$ & $1.19_{-0.18}^{+0.06}$      & $1.1_{-0.6}^{+0.3}$ & $1.02_{-0.18}^{+0.38}$ \\
    SF-probStoUD        &$0.217$& $0.1 - 0.3$ & $0.196_{-0.004}^{+0.010}$     & $0.2186_{-0.0103}^{+0.0018}$ & $0.1979_{-0.0066}^{+0.0016}$ \\
    SF-probQQtoQ        &$0.081$& $0.0 - 0.2$ & $0.0828_{-0.0024}^{+0.0011}$    & $0.0821_{-0.0032}^{+0.0010}$ & $0.0856_{-0.0039}^{+0.0007}$ \\
    SF-probSQtoQQ       &$0.915$& $0.8 - 1.0$ & $0.98_{-0.05}^{+0.09}$      & $0.748_{-0.029}^{+0.091}$ & $0.797_{-0.009}^{+0.004}$ \\
    SF-probQQ1toQQ0     &$0.0275$& $0.0 - 0.1$ & $0.033_{-0.011}^{+0.003}$     & $0.024_{-0.006}^{+0.008}$ & $0.023_{-0.003}^{+0.007}$ \\
    SF-etaSup           &$0.6$& $0.4 - 0.8$ & $0.644_{-0.018}^{+0.034}$     & $0.800_{-0.036}^{+0.012}$ & $0.7976_{-0.0068}^{+0.0018}$ \\
    SF-etaPrimeSup      &$0.12$& $0.0 - 0.3$ & $0.1095_{-0.0016}^{+0.0054}$    & $0.1027_{-0.0022}^{+0.0115}$ & $0.100_{-0.008}^{+0.017}$ \\
    SF-popcornRate      &$0.5$& $0.4 - 0.6$ & $0.31_{-0.05}^{+0.27}$      & $0.63_{-0.36}^{+0.12}$ & $0.51_{-0.17}^{+0.08}$ \\
    SF-mesonUDvector    &$0.5$& $0.3 - 0.7$ & $0.527_{-0.023}^{+0.027}$     & $0.459_{-0.008}^{+0.031}$ & $0.473_{-0.047}^{+0.009}$ \\
    SF-mesonSvector     &$0.55$& $0.35 - 0.75$ & $0.53_{-0.08}^{+0.07}$     & $0.55_{-0.05}^{+0.07}$ & $0.581_{-0.044}^{+0.018}$ \\
    SF-mesonCvector     &$0.88$& $0.7 - 1.1$ & $0.874_{-0.024}^{+0.021}$      & $1.10_{-0.22}^{+0.08}$ & $0.72_{-0.08}^{+0.13}$ \\
    SF-mesonBvector     &$2.2$& $2.0 - 2.4$ & $2.24_{-0.20}^{+0.16}$      & $2.34_{-0.50}^{+0.09}$ & $2.31_{-0.44}^{+0.27}$ \\
    \hline
  \end{tabular}
  \caption{Tuned \pythia{} parameters with default values, initial ranges for the tune, and the \texttt{Autotunes} result. See \citep{Sjostrand:2006za,Sjostrand:2014zea} 
  for details on the parameters. (SF=StringFlav,SZ=StringZ)}\label{tab:pythiatune}
\end{table}

\begin{table}[h]
  \renewcommand*{\arraystretch}{1.2}
  \begin{tabular}{|l|c|c|c|c|}
    \hline 
Parameter & Def. & Range & H7+Dip.+Cluster & H7+$\tilde{Q}$+ Cluster \\ \hline
alphaS &0.126234 & 0.12 -- 0.13& $0.13008_{-0.00061}^{+0.00013}$& $0.12455_{-0.00118}^{+0.00020}$\\
gConstituentMass &0.95 & 0.7 -- 1.1& $0.83_{-0.07}^{+0.16}$& $1.0045_{-0.0006}^{+0.0028}$\\
EMpTmin &1.2228 & 0.8 -- 1.4 & $1.204_{-0.033}^{+0.010}$& $1.068_{-0.020}^{+0.004}$\\
SPpTmin  &1.2228 & 0.8 -- 1.4& $1.204_{-0.033}^{+0.010}$& $1.22_{-0.74}^{+0.27}$\\
bNominalMass &4.2 & 4.0 -- 4.7& $4.76_{-0.21}^{+0.04}$& $4.2_{-0.5}^{+0.9}$\\
bConstituentMass & 5. & 4.0 -- 4.7& $4.01_{-0.13}^{+0.22}$& $4.03_{-0.15}^{+0.17}$\\
DecWt &0.62 & 0.5 -- 0.9& $0.59_{-0.05}^{+0.10}$& $0.61_{-0.04}^{+0.05}$\\
SngWt &0.74 & 0.5 -- 0.9& $0.86_{-0.11}^{+0.18}$& $0.80_{-0.37}^{+0.14}$\\
ClSmrLight &0.78 & 0.5 -- 1.0& $0.59_{-0.08}^{+0.05}$& $0.437_{-0.013}^{+0.041}$\\
ClSmrCharm &0. & 0.0 -- 0.2& $0.24_{-0.21}^{+0.04}$& $0.18_{-0.18}^{+0.04}$\\
ClSmrBottom &0.0204 & 0.0 -- 0.1 & $0.100_{-0.044}^{+0.019}$& $0.088_{-0.033}^{+0.018}$\\
ClMaxLight & 3.00254 & 3.0 -- 5.0 &$3.18_{-0.22}^{+0.11}$& $3.13_{-0.14}^{+0.08}$\\
ClMaxCharm &3.63822 & 3.0 -- 5.0& $3.34_{-0.07}^{+0.25}$& $3.68_{-0.07}^{+0.28}$\\
ClMaxBottom &3.911 & 3.0 -- 5.0& $4.4_{-0.6}^{+1.6}$& $4.9_{-2.9}^{+0.9}$\\
ClPowLight & 1.42426 & 1.0 -- 1.8 &$1.85_{-0.58}^{+0.24}$& $1.36_{-0.03}^{+0.15}$\\
ClPowCharm &2.33186 & 1.5 -- 3.0& $1.89_{-0.29}^{+1.71}$& $3.1_{-1.5}^{+0.4}$\\
ClPowBottom &0.6375 & 0.4 -- 0.8& $0.638_{-0.018}^{+0.104}$& $0.80_{-0.23}^{+0.04}$\\
PSplitLight & 0.847541 & 0.0 -- 1.5& $0.8747_{-0.0007}^{+0.0041}$& $0.935_{-0.018}^{+0.035}$\\
PSplitCharm &1.23399 & 0.0 -- 1.5& $0.637_{-0.028}^{+0.164}$& $1.20_{-0.66}^{+0.11}$\\
PSplitBottom &0.5306  & 0.0 -- 1.5& $0.599_{-0.113}^{+0.019}$& $0.69_{-0.12}^{+0.12}$\\
SingleHadronLimitCharm &0.0 & 0.0 -- 0.5& $0.0015_{-0.0015}^{+0.0156}$& $0.0012_{-0.0012}^{+0.0061}$\\
SingleHadronLimitBottom&0.0 & 0.0 -- 0.5 & $0.08_{-0.09}^{+0.12}$& $0.015_{-0.016}^{+0.009}$\\
    \hline
  \end{tabular}
  \caption{Result of the retuning of the \herwig{} event generator employing the default 
  hadronisation model (cluster model). The 22 dimensional parameter space was tuned with
   three sub-tunes and three iterations. Shown are the default values (corresponding to the
   Qtilde shower tune) as well as the parameter range and the tuning result for both showers. 
   \label{tab:herwigcluster}}
\end{table}

\section*{Acknowledgements} 
We would like to thank Leif L\"onnblad, Stefan Prestel and Holger Schulz for 
valuable discussions on the topic. 
We also thank Stefan Prestel and Malin Sj\"odahl for useful comments on the manuscript.
This work has received funding from the European Union's Horizon 2020 research and innovation programme as part of the Marie Skłodowska-Curie Innovative Training Network MCnetITN3 (grant agreement no. 722104).
This project has also received funding from the European Research Council (ERC) under the European Union’s Horizon 2020 research and innovation programme, grant agreement No 668679.

\bibliography{journal.bib}
\end{document}